\documentclass[aps,prl,twocolumn,preprintnumbers,showpacs,
superscriptaddress,nofootinbib]{revtex4}
\usepackage{lineno}
\usepackage{graphicx}
\usepackage{amsmath,amsfonts,amssymb}
\usepackage{citesort}
\usepackage{epsfig}
\usepackage[hyperindex,breaklinks]{hyperref} 
\usepackage{latexsym}
\usepackage{graphicx}
\usepackage{amsmath}
\usepackage{amsfonts}   
\usepackage{amssymb}    
\usepackage{float}
\usepackage{bm}
\usepackage{breakurl}
\usepackage{hyperref}
\usepackage{color}

\definecolor{purple}{rgb}{0.5,0,0.5}
\definecolor{dg}{rgb}{0.0, 0.5, 0.0}
\def\ka{\kappa}
\def\la{\lambda}
\def\aka{{A_{\kappa}}}
\def\ala{{A_{\lambda}}}
\def\mn{\mu\nu{\rm SSM}}

\def\n{\widetilde\chi^0}
\def\lsim{\raise0.3ex\hbox{$\;<$\kern-0.75em\raise-1.1ex\hbox{$\sim\;$}}}
\def\gsim{\raise0.3ex\hbox{$\;>$\kern-0.75em\raise-1.1ex\hbox{$\sim\;$}}}
\def    \beq            {\begin{equation}}
\def    \eeq            {\end{equation}}
\def    \bea           {\begin{eqnarray}}
\def    \eea           {\end{eqnarray}}

\def\nci{\nu^c_i}

\def\tb{{\rm tan}\beta}

\def\rpv{{R}_{p} \hspace{-0.37cm}\slash\hspace{0.2cm}}
\def\MET{{E}_{\rm T} \hspace{-0.38cm}\slash\hspace{0.2cm}}
\def\nbg{{\ni} \hspace{-0.20cm}\slash\hspace{0.1cm}}

\def\n{\widetilde\chi^0}
\def\c{\widetilde\chi^\pm}
\def\C{\widetilde\chi^\mp}

\def\w{W^\pm}
\def\S{S^0}
\def\P{P^0}
\begin{document}

\title{Hunting physics beyond the standard model with unusual
$\w$ and $Z$ decays}

\author{Pradipta~Ghosh}
\email{pradipta.ghosh@uam.es} 
\affiliation{Departamento de F\'{\i}sica Te\'{o}rica, Universidad Aut\'{o}noma de Madrid,
Cantoblanco, 28049 Madrid, Spain}
\affiliation{Instituto de F\'{\i}sica Te\'{o}rica UAM--CSIC, Campus de Cantoblanco, 
28049 Madrid, Spain}

\author{Daniel~E.~L\'opez-Fogliani}
\email{daniel.lopez@df.uba.ar}
\affiliation{Departamento de F\'{\i}sica, Universidad de Buenos Aires 
\& IFIBA-CONICET, 1428 Buenos Aires, Argentina}

\author{Vasiliki~A.~Mitsou}
\email{vasiliki.mitsou@ific.uv.es}
\affiliation{Instituto de F\'{\i}sica Corpuscular CSIC--UV, c/ Catedr\'atico 
Jos\'e Beltr\'an 2, 46980 Paterna (Valencia), Spain}

\author{Carlos~Mu\~noz} 
\email{carlos.munnoz@uam.es}
\affiliation{Departamento de F\'{\i}sica Te\'{o}rica, Universidad Aut\'{o}noma de Madrid,
Cantoblanco, 28049 Madrid, Spain}
\affiliation{Instituto de F\'{\i}sica Te\'{o}rica UAM--CSIC, Campus de Cantoblanco, 
28049 Madrid, Spain}

\author{Roberto~Ruiz~de~Austri}
\email{rruiz@ific.uv.es}
\affiliation{Instituto de F\'{\i}sica Corpuscular CSIC--UV, c/ Catedr\'atico 
Jos\'e Beltr\'an 2, 46980 Paterna (Valencia), Spain}

\date{\today}

\begin{abstract}
Nonstandard on-shell decays of $\w$ and $Z$ bosons are possible
within the framework of extended supersymmetric models,
i.e., with singlet states and/or new couplings compared to 
the minimal supersymmetric standard model. These
modes are typically encountered in regions of the 
parameter space with light singlet-like scalars, pseudoscalars, 
and neutralinos. 
In this letter we emphasize how these states 
can lead to novel signals at colliders from $Z$- or 
$\w$-boson decays with prompt or displaced 
multileptons/tau jets/jets/photons in the final states.
These new modes would give distinct evidence of new physics 
even when direct searches remain unsuccessful. 
We discuss the possibilities of probing these new
signals using the existing LHC run-I data set. 
We also address the same in the context of the
LHC run-II, as well as for the future colliders.
We exemplify our observations with the ``$\mu$ from $\nu$'' supersymmetric 
standard model, where three generations of right-handed
neutrino superfields are used to solve shortcomings of the minimal
supersymmetric standard model. 
We also extend our discussion for other variants of supersymmetric 
models that can accommodate similar signatures.

\end{abstract}
\preprint{
FTUAM-14-8,
~~~IFT-UAM/CSIC-14-020,
~~~IFIC/14-19}


\maketitle

%
\section*{I. INTRODUCTION} 

\vspace*{-0.35cm}

Physics beyond the standard model (SM) remains
necessary even after the long-awaited discovery of the Higgs boson \cite{Higgs}.
A much anticipated but hitherto unseen excess over the SM
thus makes it rather essential and timely to explore other methods, 
e.g. precision measurements of the SM observables, where evidence 
of new physics may remain hidden.
In this article we present an analysis of this kind
in the context of extended supersymmetric (SUSY) models,
which can lead to new two-body decays of $\w$ and $Z$ bosons
for certain regions of the \textit{enlarged} parameter space.
Concerning a $Z$ boson, decaying to a scalar and a pseudoscalar, 
we search for final states with a combination of
four prompt leptons $(\ell=e,\,\mu$)/$\tau$ jets (from hadronically 
decaying $\tau$)/jets/photons. 
We look for similar but displaced final states with missing transverse 
energy $(\MET)$ when a $Z$ boson decays into a pair of neutralinos. 
For the $\w$ boson, when it decays into a charged
lepton and a neutralino, we look for final states with two displaced
leptons/$\tau$ jets/jets/photons + $\MET$ together with
a prompt lepton/$\tau$ jet.

Earlier analyses of unusual $\w$ decays~\cite{Wd}
in the minimal supersymmetric standard model (MSSM) 
are already excluded by the chargino mass bound 
\cite{PDG}. Unusual $Z$ decays for models with an extended Higgs
sector with/without SUSY have also been studied,
both theoretically \cite{rpcZ} and experimentally \cite{Zexp}.
These decays lead to either multiparticles
or missing transverse momentum/energy signatures 
at colliders. Most of these scenarios
e.g., a light $(<{M_Z}/{2})$ doublet-like pseudoscalar
as discussed in the third and sixth papers of Ref. \cite{rpcZ}
hardly survive with the current experimental constraints.
However, a study of all new $\w$ and $Z$ 
two-body decays through singlet-like 
states in the light of Higgs boson discovery \cite{Higgs} is 
missing to date. 
In this article we aim to carry out this 
analysis using as a case study the 
``$\mu$ from $\nu$'' supersymmetric standard model
($\mn$) \cite{mn1,mn2},
the \textit{simplest} variant of the 
MSSM to house these signals, apart from
housing nonzero neutrino masses and mixing \cite{mn1,mn2,PG1,Hirsch2009,mnSCPV,PG2}
and offering a solution to the $\mu$ problem \cite{muprob}
of the MSSM. Nonetheless, we also extend our discussion
to other model variants. Note that the said light singlet-like states 
also contribute to the invisible/nonstandard decay
branching fractions (BRs) for a Higgs-like scalar \cite{hns,hns1}.

\vspace*{-0.15cm}

\section*{II. THE {\boldmath$\mu\nu$}SSM} 
\vspace*{-0.25cm}

In the $\mn$, three families of
right-handed neutrino superfields ($\hat\nu^c_i$) 
are instrumental in offering a solution to the 
$\mu$ problem \cite{muprob} of the MSSM,
and concurrently in housing the observed pattern of
neutrino masses and mixing \cite{mn1,mn2,PG1,Hirsch2009,mnSCPV,PG2}.
The superpotential is given by
%
{\small
\bea\label{superpotential}
W  &= 
\ \epsilon_{ab} (
Y_{u_{ij}} \, \hat H_u^b\, \hat Q^a_i \, \hat u_j^c +
Y_{d_{ij}} \, \hat H_d^a\, \hat Q^b_i \, \hat d_j^c +
Y_{e_{ij}} \, \hat H_d^a\, \hat L^b_i \, \hat e_j^c 
\nonumber\\ 
&
+ Y_{\nu_{ij}} \, \hat H_u^b\, \hat L^a_i \, \hat \nu^c_j -   
\lambda_{i} \, \hat \nu^c_i\,\hat H_d^a \hat H_u^b)+
\frac{1}{3}
\kappa{_{ijk}} 
\hat \nu^c_i\hat \nu^c_j\hat \nu^c_k.
\eea}
%
The last three terms in Eq.~(\ref{superpotential}) break $R$ parity
($\rpv$) explicitly.
After the spontaneous breaking of the electroweak symmetry,
the neutral scalars develop vacuum expectation values 
as {\small $\langle H_d^0 \rangle = v_d,\,\langle H_u^0 \rangle = 
v_u,\,\langle \tilde \nu_i \rangle = \nu_i,\, {\rm and~} \langle \tilde \nu_i^c 
\rangle = \nu_i^c$}; thus, they generate the effective bilinear terms 
$\varepsilon_i\,\hat L_i \hat H_u$,
$\mu\,\hat H_d \hat H_u$ and Majorana mass terms $M_{{ij}}\hat \nu^c_i \hat\nu^c_j$,
with $\varepsilon_i\equiv Y_{\nu_{ij}}\nu^c_j$,
$\mu\equiv \lambda_i\nu^c_i$, and $M_{{ij}}\equiv 2\kappa_{ijk} \nu^c_k$.

The enlarged field content and the $\rpv$ in the $\mn$
result in eight $CP$-even $(\S_\alpha)$ and seven $CP$-odd $(\P_\alpha)$ neutral 
scalar, seven charged scalar $(S^\pm_\alpha)$, ten neutralino $(\n_\alpha)$,  
and five chargino $(\c_\alpha)$ states \cite{mn1,mn2,PG1,PG2}.
The three lightest neutralinos and charginos,
denoted as $\n_i$ and $\widetilde\chi^\pm_i$,
coincide with the left-handed neutrinos and 
the charged leptons.

In the $\mn$, light (in order to trigger new $Z,\,\w$, and Higgs 
decays, i.e., $\lsim M_Z/2$ \cite{Hirsch2009,PG3,mnLHC1,PG4,PG5}) 
right-handed sneutrino ($\widetilde\nu^c$) 
and right-handed neutrino $(\nu_R)$-like states,
in the bottom of the mass spectrum,
are possible with suitable choices of   
$\la_i$, $\ka_{ijk}$, $\tb~(=\frac{v_u}{v_d})$, 
$\nu^c_i$, and the soft SUSY-breaking parameters \cite{mn1,mn2} 
$A_{\kappa_{ijk}}$ and $A_{\la_i}$. The parameters
$\la_i$ and $A_{\la_i}$ control the doublet impurity 
and hence affect the lightness of these states \cite{mnLHC1,PG4,PG5}. 
The parameters $\ka_{ijk}$, $A_{\ka_{ijk}}$, and
$\nu^c_i$ determine their mass scales \cite{PG4,PG5}.
A small doublet component, i.e. a small $\la_i$,
together with small $\tb$ values makes it easier (see, e.g.,
Ref.~\cite{small-mixing}) for 
these states to evade a class of collider 
\cite{collider-LEP,collider-tev,collider-LHC} and 
low-energy constraints \cite{Upsilon,Bsmumu,mug-2} 
(see Ref.~\cite{PG5} for details).
Here, $\S_i,\,\P_i$, and $\n_{i+3}$ are used
to denote $\widetilde \nu^c$-like scalars, pseudoscalars,
and $\nu_R$-like neutralinos in the mass eigenbasis, respectively.
With this notation, $S^0_4$ is 
the SM-like Higgs boson \cite{mn2,PG3,mnLHC1,PG4,PG5}.
The effects of $\S_i$, $\P_i$, and $\n_{i+3}$ states in $\S_4$ decays 
have already been addressed in the $\mn$ before \cite{PG3,mnLHC1} 
and after \cite{PG4,PG5} the Higgs discovery \cite{Higgs}.

\begin{figure}[ht]
\includegraphics[height=5.5cm,width=8.500cm]{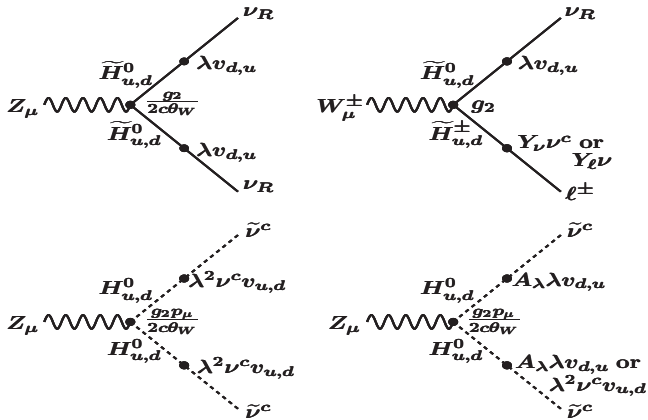}
\caption{Leading on-shell $\w$ and $Z$ decays
through $\nu_R$ and $\widetilde\nu^c$, where 
$g_2$ is the SU(2) gauge coupling, $\theta_W$ is the Weinberg angle,
${\rm c}\equiv$ cos and $p_\mu$ is the momentum factor
for the $Z\S_i\P_j$ vertices. Family indices and 
extra factors coming from the field decomposition
(as mentioned in the text) are not explicitly shown.}
\label{fig:wzdecay}
\end{figure}

\vspace*{-0.15cm}

\section*{III. NEW {\boldmath$Z,\w$} DECAYS IN THE 
{\boldmath$\mu\nu$}SSM} 
\vspace*{-0.25cm}

New on-shell decays, like $Z$ $\to$ $\S_i \P_j$, $\n_{i+3}\n_{j+3}$ 
and $\w$ $\to$ $\c_i \n_{j+3}$, as already stated, naturally open up
for $m_{\S_i,\,\P_i,\,\n_{i+3}} \lsim M_Z/2$. 
These decays are schematically shown in Fig.~\ref{fig:wzdecay}
using the flavor basis. Depending on $m_{\S_i,\,\P_i}$ \cite{PG5}, 
successive $\S_i, \P_i$ decays into a pair of
leptons/$\tau$ jets/jets/photons give prompt multiparticle
final states for the $Z\to\S_i\P_j$ processes. 
A light neutralino (with a mass $<20$ GeV) in SUSY models
with singlets can decay through lighter\cite{Hirsch2009}/slightly 
heavier \cite{PG4} $\S_i/\P_i+\n_j$ modes.
These new modes dominate for $m_{\n}<40$ GeV and 
can give displaced decays within the tracker \cite{Hirsch2009,mnLHC1,PG4}.
The subsequent $\S_i/\P_i$ decay gives a combination of four displaced but detectable 
leptons/$\tau$ jets/jets/photons \cite{PG4,PG5} 
for the $Z\to\n_{i+3}\n_{j+3}$ modes, 
with some $\MET$ from neutrinos and possible mismeasurements. 
Note that the $\S_4\to \S_i\S_j$, $\P_i\P_j$, and $\n_{i+3}\n_{j+3}$ 
decays \cite{PG5} produce final states that are similar to 
those of $Z\to \S_i\P_j,\,\n_{i+3}\n_{j+3}$ processes.
The $\w\to\c_i\n_{j+3}$ decays, in the same way, lead to
final states with one prompt lepton/$\tau$ jet + two 
displaced leptons/$\tau$ jets/jets/photons
+ $\MET$. 
Possible final states are shown below in tabular format (see 
Table \ref{Tab:signals}) with $x=$ lepton/$\tau$ jet/jet/photon. 
The superscript $P$ and $D$ are used to denote prompt and displaced nature,
respectively.

\begin{table}[h]
\footnotesize
\caption{\label{Tab:signals}Table showing all possible final states
from nonstandard $Z$ and $\w$ decays along with their respective origins.}
\begin{ruledtabular}
\begin{tabular}{ c  c  c  }
  & $Z$ decay & $\w$ decay  \\ \hline
&  $2 x^D 2 {\bar{x}}^D$ + $\MET$ $({\rm via~} \n_{i+3}\n_{j+3})$
& $\ell^P$/$\tau$-jet$^P$ 
+ $x^D {\bar{x}}^D$ \\
& $2 x^P 2 {\bar{x}}^P $ $({\rm via~} S^0_iP^0_j)$ &
+ $\MET$ $({\rm via~} \widetilde\chi^\pm_i\n_{j+3})$
\end{tabular}
\end{ruledtabular}
\end{table}

%
These new decays are constrained by  
the measured decay widths, i.e., $\Gamma_Z=$ $2.4952 \pm 0.0023$ GeV,
$\Gamma_{\w}=$ $2.085 \pm 0.042$ GeV, and 
$\Gamma^{\rm inv}_{Z}=$ $0.499 \pm 0.0015$ GeV \cite{PDG}
if $\n_{i+3}$ decays invisibly, i.e., outside the 
detector. 
From the latest theoretical calculation, including
higher-order contributions, one gets 
$\Gamma^{\rm theo}_Z= 2.49424 \pm 0.0005$ GeV  
and $\Gamma^{\rm inv(theo)}_Z= 0.50147 \pm 0.000014$ GeV 
at the $1\sigma$ level \cite{Zwdth}. So one can compute $\delta \Gamma_Z\equiv
\Gamma_Z - \Gamma^{\rm theo}_Z = 1.0 \pm 2.4$ MeV
and, similarly, $\delta \Gamma^{\rm inv}_Z = -2.5 \pm 1.5$ MeV,
where we have added the theoretical and experimental $1\sigma$
errors in quadrature. It is thus apparent that at the 
$1\sigma$ level\footnote{We restrict ourselves within $68\%$ C.L. 
of errors (see $11${th} paper of Ref.~\cite{rpcZ}), 
which gives more stringent constraints.} one has a freedom
of about $3.4$ MeV to accommodate the new physics
contributions in $Z$ decays over the SM prediction. Concerning
new invisible $Z$ decays beyond the SM, one is forced to
consider a $2\sigma$ variation in order to accommodate
a positive new physics effect with a freedom of about $0.5$ MeV.
A similar analysis for $\w$ with $\Gamma^{\rm theo}_{\w}=2.0932 \pm 0.0022$
GeV \cite{Wwdth} gives $\delta \Gamma_{\w} = -8 \pm 42$ MeV.
Thus, any new contribution must give a decay width
less than about $34$ MeV at the $1\sigma$ level.
Moreover, from the SM one gets  
BR$(Z\to 4\ell^P)$ $= 4.2^{+0.9}_{-0.8}\times 10^{-6}$ \cite{PDG,Zto4l}
and BR$(Z\to 4b^P)$ $=3.6^{+1.3}_{-1.3}\times 10^{-4}$ \cite{PDG,Zto4b}.
The latest BR$(Z\to 4\ell^P)$ is 
estimated as $3.2^{+0.28}_{-0.28}\times 10^{-6}$ \cite{smz4ln2}.
Regarding final states, $\S_i,\P_i\to b\bar{b}$ remains
generic for $2m_b$ $\lsim m_{\S_i,\,\P_i}$ $\lsim M_Z/2$ 
while taus dominate for 
$2m_\tau\lsim$ $m_{\S_i,\,\P_i}$ $\lsim 2m_b$ \cite{mnLHC1,PG4,PG5}.
Light jets $(\nbg b)$ or leptons normally
emerge for $m_{\S_i,\P_i}\lsim 2m_\tau$ with a
moderate parameter tuning, especially for $\la_i,\,\kappa_{ijk}$,
$\nu^c_i$, and ${A_\kappa}_{ijk}$ \cite{PG5}. 
A multiphoton signal through the $\S_i/\P_i\to \gamma\gamma$ process 
(that is \textit{unique} in nature) requires large parameter tuning to 
get a statistically significant result. Note that a
low mass of the mother particle, e.g. $M_Z\approx 91$ GeV
or $M_W\approx 80$ GeV, typically
gives rise to soft leptons/$\tau$ jets/jets/photons, 
in the final states~\cite{PG4,PG5}, 
i.e. with low transverse momentum, 
that suffer from poor detection efficiency \cite{tdr-btau}.

Concerning the lightest neutralino $\n_4$, a
Higgsino- or wino-like nature is forbidden 
for $2m_{\n_4}\lsim M_Z$ from the lighter chargino mass bound \cite{PDG}. 
The latter demands the minimum of $(\mu,M_2)$ $\gsim 100$ GeV,
where $M_2$ is the SU(2) gaugino soft-mass.
Further, the tree-level $Z$-Higgsino-Higgsino 
interaction constrains the amount of Higgsino impurity in $\n_4$
from the measured $\Gamma_Z$ \cite{PDG}.
So, depending on the relative orders of $2\ka_{ijk}\nu^c_k$
and $M_1$ [U(1) gaugino soft-mass], a light $\n_4$ 
is either bino- or right-handed neutrino-like, or 
a bino-right-handed neutrino mixed state.
A light bino-like $\n_4$ needs a low $M_1$ 
and hence a relaxation of the gaugino mass unification 
at the high scale when considered together with the LHC bound 
on gluino mass \cite{LHClim}. 
Besides, as lighter $\S_i,\,\P_i$ states
are not assured for a bino-like $\n_4$ \cite{PG5}, 
typical decay occurs beyond the tracker 
and often outside the detector~\cite{Hirsch2009,mnLHC1}.
Hence we do not consider this possibility.
One can also consider a bino-like $\n_7$, with $2m_{\n_7}\lsim M_Z$,
that decays to right-handed neutrino-like $\n_{i+3}$ + $\S_j/\P_j$, followed by 
$\n_{i+3}\to \n_j + \S_k/\P_k$, and finally $\S_i/\P_i \to$
a pair of leptons/$\tau$ jets/jets/photons.
In this scenario, in spite of the large particle
multiplicity of the final state, most of these leptons/$\tau$ jets/jets/photons
remain undetected due to their soft nature 
\cite{PG5}, and thus is not studied here.

{\section*{IV. NEW {\boldmath$Z,\w$} DECAYS IN OTHER MODELS}} 
\vspace*{-0.15cm}

Unusual $Z$ decays with prompt final states also appear in 
$R_p$-conserving models with singlets, e.g., the next-to-MSSM (NMSSM) 
(see fifth paper of Ref.~\cite{small-mixing} for a review). The presence
of $3\hat\nu^c_i$ in the $\mn$, which emerges naturally from the family
symmetry, can produce different peaks in the 
invariant mass $m_{\rm inv}/{\rm M_{T2}}$ \cite{mt2}
distributions for the two-leptons/$\tau$ jets/jets/photons systems. 
In the NMSSM, however, one expects
a single peak with one singlet superfield. 
With dedicated search strategies, e.g., better detection efficiency
for a soft lepton/$\tau$ jet/jet/photon, etc., and higher statistics 
one could hope to probe (and hence discriminate) these scenarios in the coming years.
The model of Ref.~\cite{kitano} in the context of the 
studied signals is similar to the $\mn$, although additional constraints
can appear for the former from dark matter searches.
Note that in SUSY models with singlets, 
$\S_4\to \S_i\S_j$, $\P_i\P_j$, 
and $\n_{i+3}\n_{j+3}$ processes \cite{mnLHC1,PG5,PGD}
can mimic new $Z$ decays, but they
give a different $m_{\rm inv}/M_{\rm T2}$ peak for the four-particle 
system around $m_{\S_4}$.

Displaced $\w$ and $Z$ decays 
are also possible in the MSSM with $\rpv$ \cite{rpvZ},
sometimes with a richer topology, e.g. two-leptons/$\tau$ jets 
+ 4 jets +$\MET$
with $\la'_{ijk}\hat L_i \hat Q_j \hat d^c_k$ couplings.
For the bilinear $\rpv$ model, a light neutralino with a mass $<20$ GeV 
normally decays outside the detector \cite{Hirsch2009}.
The MSSM with trilinear $\rpv$ couplings 
[$\la_{ijk}\hat L_i \hat L_j \hat e^c_k$ for 
$\la_{ijk}\sim {\cal{O}} (10^{-3})$], however,
can produce a decay length within $1~{\rm cm}$-$3{\rm m}$ \cite{trpv} 
for a neutralino in the same mass range.
Nevertheless, with different intermediate states it
is possible to identify (and hence discriminate) these signals 
by constructing a set of kinematical variables.
However, one needs a good reconstruction efficiency
for the displaced and normally soft lepton/$\tau$ jet/jet/photon
to probe these signals.

Displaced $Z$ decays in the NMSSM can appear 
for fine-tuned $\la$ values \cite{nmssm1},
when a pair of next-to-lightest SUSY particles (NLSPs) 
are produced in the $Z$ decay. 
These NLSPs, when decay further into a lightest supersymmetric
particle and a scalar/pseudoscalar, followed by the scalar/pseudoscalar 
decays into two leptons/$\tau$ jets/jets/photons
produce identical displaced final states. The corresponding
decay length, however, is never in the range of $1{\rm cm}-3{\rm m}$ 
(second paper of Ref.~\cite{nmssm1}).

Look-alike displaced states also appear for the model
of Ref.~\cite{kitano} even with an
$R_p$-conserving vacuum, when a $Z$ decays into 
two singlino-like NLSPs, followed by an
NLSP $\to$ $\widetilde\nu^c$-like lightest supersymmetric
particle and a right-handed neutrino process. 
This right-handed neutrino then decays further to 
a scalar/pseudoscalar and a left-handed neutrino, giving 
an identical signal. However, in this case $\MET$ 
could be larger and the scenario is 
constrained from dark matter searches.

These observations verify the $\mn$ as the \textit{minimal
extension} beyond the MSSM to house these distinctive 
collider signatures together with the correct neutrino physics
as well as offering a solution to the $\mu$ problem.

\vspace*{0.35cm}

{\section*{V. THE BACKGROUNDS}}

Leading SM backgrounds for $Z$ $\to$ $\S_i\P_j$ 
processes come through Drell-Yan,  
$\w W^\mp$, $\w Z$, $ZZ/\gamma$, $b\bar{b}$,
dileptonically decaying $t\bar{t}$, $\w/Z + {\rm jets}$ 
and $Z\to 4l$, $4~{\rm jets}$, $2l2~{\rm jets}$, where
$l$ represents a charged lepton. A
faithful reconstruction of $m_{\rm inv}$/${\rm M}_{\rm T2}$
for the $4l$/$4~{\rm jets}$/$2l2~{\rm jets}$ 
as well as $2l/2~{\rm jets}$ systems, with a proper 
experimental setup (as mentioned earlier) can 
disentangle these signals. 
The latter is crucial to isolate the studied
signals from $Z\to$ $4l$, $4~{\rm jets}$, $2l2~{\rm jets}$
backgrounds \cite{smz4lo}, where the $m_{\rm inv}$/$M_{\rm T2}$ 
distribution for four-particle systems
also peaks around $M_Z$. A recent analysis \cite{smz4ln2}
only considered $e$ and $\mu$, while 
in the $\mn$ a $\tau(b)$ rich signal is generic for 
$2m_\tau \lsim m_{\S_i,\,\P_i}\lsim 2m_b~(m_{\S_i,\,\P_i}\gsim 2m_b)$.
Displaced $Z$, $W^\pm$ decays are exempted from the SM backgrounds.

\vspace*{0.35cm}

{\section*{VI. ESTIMATING NEW BRs AND DECAY WIDTHS}}

Turning back to the $\mn$, we aim to estimate BRs for 
nonstandard $\w,Z$ decays analytically,
and hence we stick to the flavor basis for an easy
interpretation. $Z$ decays - which are suppressed 
by small $Y_{\nu_{ij}}/\nu_i$, as required 
for a TeV-scale seesaw \cite{mn1,mn2,PG1,PG2,Hirsch2009,mnSCPV} -
are not shown in Fig.~\ref{fig:wzdecay}. 
The smallness of $Y_{\nu_{ij}}$ and $\nu_i$ assure
that BRs for the tree-level flavor-violating $Z\to\c_i\C_j$ 
processes are well below the 
respective SM limits. Note that the same
logic also predicts very small BRs
for flavor-violating Higgs decays into a pair
of leptons. Hence, it remains difficult to
accommodate the recent excess in the $\mu\tau$
final state from Higgs decays, as reported by the CMS
Collaboration \cite{HiggsLFV}.
For $\w$, on the contrary, processes
involving $Y_{\nu_{ij}}/\nu_i$ are the 
leading nonstandard decay modes, and hence
always severely suppressed by orders of magnitude 
compared to the unusual $Z$ decays.

We start our discussion with the complete expressions for 
the new $\w$ and $Z$ decay widths, i.e.,
$\Gamma(\w\to \c_i\n_{j+3})$, 
$\Gamma(Z\to \n_{i+3}\n_{j+3})$, and $\Gamma(Z\to \S_i\P_j)$.
These are written as follows:

\begin{widetext}
{\small
\bea\label{decaywidth}
&&\Gamma(\w \to \c_i\n_{j+3})
= \frac{g^2_2 }{48\pi M^5_W }
\,\left[\left(M^2_W-m^2_{\c_i}-m^2_{\n_{j+3}}\right)^2
-4 m^2_{\c_i}m^2_{\n_{j+3}}\right]^{\frac{1}{2}}\times 
\left[\left(\left|O^{cnw}_{Lij+3}\right|^2
+\left|O^{cnw}_{Rij+3}\right|^2\right)\right. \nonumber\\
&&\left.\times\left(2 M^4_W-M^2_W\left(m^2_{\c_i}+m^2_{\n_{j+3}}\right)-
\left(m^2_{\c_i} - m^2_{\n_{j+3}}\right)^2\right)
+ 12 \Re\left(O^{cnw^*}_{Lij+3}
O^{cnw}_{Rij+3}\right) m_{\c_i}m_{\n_{j+3}} M^2_W \right],\nonumber\\
&&\Gamma(Z \to \n_{i+3}\n_{j+3})
= \frac{g^2_2 }{48\pi M^5_Z}
\,\left[\left(M^2_Z-m^2_{\n_{i+3}}-m^2_{\n_{j+3}}\right)^2
-4 m^2_{\n_{i+3}}m^2_{\n_{j+3}}\right]^\frac{1}{2} 
\times\left[\left(\left|O^{nnz}_{Li+3j+3}\right|^2
+\left|O^{nnz}_{Ri+3j+3}\right|^2\right)\right.\nonumber\\
&&\left.\times\left(2M^4_Z-M^2_Z\left(m^2_{\n_{i+3}}+m^2_{\n_{j+3}}\right)-
\left(m^2_{\n_{i+3}} - m^2_{\n_{j+3}}\right)^2\right)
+12 \Re\left(O^{nnz^*}_{Li+3j+3}
O^{nnz}_{Ri+3j+3}\right) m_{\n_{i+3}}m_{\n_{j+3}} M^2_Z \right],\\
&&\Gamma(Z\to \S_i\P_j)=\frac{g^2_2 \left|O^{spz}_{ij}\right|^2}
{192\pi{\rm cos^2\theta_W} M^5_Z}\,
\left[\left(M^2_Z-m^2_{\S_i}-m^2_{\P_j}\right)^2
-4 m^2_{\S_i}m^2_{\P_j}\right]^\frac{1}{2}\times
\left[M^4_Z-2M^2_Z\left(m^2_{\S_i}+m^2_{\P_j}\right)
+\left(m^2_{\S_i}-m^2_{\P_j}\right)^2\right].\notag
\eea}
\end{widetext}
The couplings $O^{nnz}_{L(R)}$, $O^{cnw}_{L(R)}$
are given in Ref. \cite{PG2}, and
$O^{spz}_{ij}= \{R^{\S}_{i1}R^{\P}_{j1}
-R^{\S}_{i2}R^{\P}_{j2} + \sum^{3}_{p=1} R^{\S}_{i,p+5}R^{\P}_{j,p+5}\}$,
where $R^{S^0}_{ab}$ $(R^{\P}_{ab})$ denotes the 
amount of the bth flavor state in the ath
scalar (pseudoscalar) mass eigenstate
after rotating away the Goldstone boson.
We will use these full formulas 
in a forthcoming publication \cite{PgF}
for a complete numerical analysis of
the unusual $\w,\,Z$ decays over the different
regions of the parameter space. Thus, 
in this article we will use a set of
approximate formulas for various decay widths 
to explore the behavior of these new decays
analytically.

The relative strength of the three $Z\to \S_i\P_j$ processes in 
Fig.~\ref{fig:wzdecay}, assuming $\nu^c\approx \ala$,
goes as $\la^2:1:\la$. Clearly, the process $\propto A^2_\la$
dominates over the rest when $\la$ is small $(\lsim 0.1)$,
unless  $\nu^c\gsim 10\ala$.
A small $\la$ also ensures singlet purity for the light $\S_i,\,\P_i$, 
and $\n_{i+3}$ states \cite{mn2,Hirsch2009,PG3,mnLHC1,PG4,PG5}.
Note that the decay BR of a light $\S_i,\P_i$, 
or $\n_{i+3}$ state into a specific mode
can be tuned to $100\%$ depending on $m_{\S_i,\P_i}$ \cite{mnLHC1,PG5}.
In this derivation and for the subsequent analysis we assume, for simplicity, 
universal $\la_i$, $\ala_i$, $\nci$,
i.e., ${\mu=3\lambda\nu^c}$ and
$(\ka,\,\aka)_{ijk}=(\ka,\,\aka)_i\delta_{ij}\delta_{jk}$
with $\ka_i \approx \ka_j$ \cite{mnLHC1,PG4}.
Further, we take $(Y_\nu,\,A_\nu$\cite{mn1,mn2}$)_{ij}
=(Y_\nu,\,A_\nu)_{i}\delta_{ij}$. These approximate
formulas are also valid in the mass basis, provided that
the doublet contamination in $\S_i$ and $\P_i$ states
is negligible.

$\Gamma(Z\to\n_{i+3}\n_{j+3})$
and $\Gamma(Z\to\S_i\P_j)$, following their orders in 
Fig.\ref{fig:wzdecay}, are estimated as 
{\small
\bea\label{Z-approx}
\Gamma_{Z\n\n} &\approx& \frac{6g^2_2\la^4v^4_{u}M^2_Z}{2^6 c^2_{\theta_W}
\mu^4}
{\mathcal{P}},\,
\Gamma^1_{Z\S\P}\approx 
\frac{6g^2_2 \la^4 \mu^4  v^4_{u}M^2_Z}{2^6 3^4c^2_{\theta_W}\widetilde m^8}
\mathcal{P},\\
\Gamma^2_{Z\S\P}&\approx& 
\frac{6g^2_2 A^4_\la \la^4 v^4_{u} M^2_Z}{4c^2_{\theta_W}\widetilde m^8}
\mathcal{P},\,
\Gamma^3_{Z\S\P}\approx 
\frac{6g^2_2 A^2_\la \la^4 {\mu}^2 v^4_{u}M^2_Z}{2^4 3^2c^2_{\theta_W}\widetilde m^8}
\mathcal{P},\notag
\eea}
where $\mathcal{P}=\frac{1}{16\pi m_A}\sqrt{\left\{1-\left(\frac{m^2_B}{m^2_A}
+\frac{m^2_c}{m^2_A}\right)\right\}^2-4\frac{m^2_B}{m^2_A}\frac{m^2_C}{m^2_A}}$
is the phase-space factor for a $A\to B C$ process.
A factor of $``6$'' in the numerator comes after summing 
over all possible $i$ and $j$ values without double 
counting.\footnote{One can write a set of similar
formulas for $``n$'' families of right-handed
neutrino superfields when the numerator looks like
$(\frac{n+1}{2n})$ {\boldmath $\la^4$}. Here {\boldmath $\la^2$}
$\equiv \sum_i \la^2_i = n\la^2$, assuming universal $\la_i$. 
The quantity {\boldmath $\la^2$} is bounded from above
from the requirement of maintaining the perturbative nature
of $\la_i$ parameters up to some energy scale, 
e.g., $\lsim (0.7)^2$ when the scale lies at $10^{16}$ GeV \cite{mn2}. Hence,
adding more and more singlets, i.e., $n\to \infty$ does not imply
a blow-up behavior for the new $Z$-decay widths.}
We have also introduced a factor 
(not shown in Fig. \ref{fig:wzdecay}) coming from the field
decomposition as discussed in Ref. \cite{PG5} following
the expressions given in Ref. \cite{mn2}. Those factors are
$2^4,\,2^4,\,1$, and $2^2$, respectively for the four
decay widths shown in Eq. (\ref{Z-approx}).
The generic mass scale for the intermediate Higgsino (Higgs) 
is denoted by $\mu~(\widetilde m)$. For $Z$ decays at rest, $p_\mu 
p^\mu=M^2_Z$.

In order to estimate the maximum new $\Gamma_Z$ we
have neglected contributions $\propto v_d$ 
compared to $v_u$ for $\tb>1$,
and thus we have used $v_u \approx v = 174$ GeV. 
Now with $g_2=0.652$, $\la=0.1$,
${\rm c}^2_W=0.769$, and $M_Z=91.187$ GeV \cite{PDG}, and 
assuming $\mu,\,A_\la,\,\widetilde m $ $\approx {\cal{O}}(v)$,
the total new $\Gamma_Z$ (from three leading 
contributions, i.e., $\Gamma_{Z\n\n}$ + $\Gamma^2_{Z\S\P}$ + $\Gamma^3_{Z\S\P}$)
from Eq.~(\ref{Z-approx}) is evaluated as $\approx 0.16$ MeV.
This number is smaller than $3.4$ MeV, as estimated in Sec. III.
The total new $\Gamma_Z$ in this region
of parameter space for $\la\gsim 0.21$ becomes 
larger than $3.4$ MeV and hence experimentally disfavored. 
Further, we also note that the lowest possible value
for the $\mu$ parameter is $\approx 100$ GeV, as required from
the lighter chargino mass bound \cite{PDG}. In this 
region of parameter space, i.e., when 
$\mu\approx 100$ GeV while $\widetilde m, \, 
A_\lambda \approx {\mathcal{O}}(v)$, the total new $\Gamma_Z$
is larger than $3.4$ MeV for $\la \gsim 0.19$.
In this estimation we assume $m_{\S_i}$ $\sim$ $m_{\P_i}$ $\sim$ $m_{\n_{i+3}}$
and consider $\mathcal{P}\approx 1/16\pi M_Z$. This
approximation remains valid as long as the $\S_i,\,\P_i$,
and $\n_{i+3}$ states are much lighter than $M_Z$. 
For heavier $\S_i,\,\P_i$, and $\n_{i+3}$ states,
the new $\Gamma_Z$ reduces due to a suppression from the
phase-space factor.
We note in passing that in this analysis
we focus only on ``visible'' $\n_{i+3}$ decays.
The corner of the parameter space (in the 
case of genuinely invisible $Z$ decays) with 
 $\mu\lsim 130$ GeV lowers the upper limit of $\la$
(e.g., $\lsim 0.16$ using $\mu=100$ GeV)
from a freedom of $0.5$ MeV in the measurement
of the invisible $Z$-decay width. For a larger value
of the $\mu$ parameter, $\Gamma_{Z\n\n}$ appears
to be quite suppressed compared to $\Gamma^2_{Z\S\P}$ 
[see Eq.~(\ref{Z-approx})], and hence the constraint
on the total $Z$-decay width is reached faster than that
of the invisible $Z$ decay. The latter 
observation reverses for $\widetilde m \gsim 1.3\mu$
with $\ala \approx \widetilde m$ irrespective of the 
magnitude of $\ala,\,\mu$, and $\widetilde m$ with respect
to the scale of $v$.

One can, nevertheless, use $\ala,\,\widetilde m,\,\mu$ $> v$
to suppress the new $\Gamma_Z$ as well as the
new $\Gamma^{\rm inv}_Z$ for any $\la$ values. This
way, larger $\la$ values can survive the experimental constraint on $\Gamma_Z$,
e.g. $\ala,\,\widetilde m,\,\mu =2 v$ gives a
new $\Gamma_Z < 3.4$ MeV even if $\la = 0.4$.
Such large $\la$ values\footnote{Note that by
assuming the perturbative nature of $\la_i$ parameters
up to the grand unified theory scale, i.e.,
$10^{16}$ GeV, one gets {\boldmath $\la^2$} $\lsim  (0.7)^2$ \cite{mn2}.
This predicts a maximum for $\la \approx 0.4$ assuming
universal $\la_i$. Higher $\la_i$ values require a lower scale
up to which the concerned parameters respect their
perturbative nature. \cite{mn2}.}, however, spoil the singlet purity
and lightness of $\S_i,\P_i$, and $\n_{i+3}$ states \cite{PG5}.

So we conclude that, for $\la\lsim 0.1$ and 
${\cal{O}}(1$ TeV) $\gsim \ala,\,\widetilde m,\,\mu \gsim $ ${\cal{O}}(v)$,
one gets ${\cal{O}}(10^{-8})\lsim$ BR$(Z\to \S_i\P_j,\,\n_{i+3}\n_{j+3}) 
\lsim {\cal{O}}(10^{-4})$.
Further, BR$(\S_i/\P_i\to x^P\bar{x}^P)$ and 
BR$(\n_{i+3}\to x^D\bar{x}^D+\MET)$ $\approx 1$
imply that BR$(Z\to 2b^P2\bar{b}^P)$ and 
BR$(Z\to 2b^D2\bar{b}^D+\MET)$ also vary
in the \textit{same} range. Thus, the new decay 
BR for the $Z\to 2b^P2\bar{b}^P$ process
(i.e., when $2m_b \lsim m_{\S_i,\,\P_i} \lsim M_Z/2$) 
remains \textit{below} the SM measured value \cite{PDG,Zto4b}, 
while $Z\to 2\ell^P2\bar{\ell}^P$ (with $m_{\S_i,\,\P_i}\lsim 2m_\tau$)
remains comparable (at the $2\sigma$ level) or below 
the concerned SM limit \cite{smz4ln2} 
for $\ala,\,\widetilde m,\,\mu \geq 356$ GeV. 
For the region of parameter space
with $\mu\approx 100$ GeV and $\ala,\,\widetilde m \approx
{\cal{O}}(v)$, one needs to consider $\la \lsim 0.044$
to respect the SM measurement of BR$(Z\to 2\ell^P2\bar{\ell}^P)$,
concerning the $2\sigma$ variation
around the central value \cite{smz4ln2}. These issues will be addressed in 
detail in Ref. \cite{PgF}.
The other way of getting $Z\to 2\ell^P2\bar{\ell}^P$ - i.e.,
through leptonic tau decays following
$Z\to2\tau2{\bar{\tau}}$ and taking 
the leptonic $\tau$ decay BR $\approx 0.35$ \cite{PDG} gives 
${\cal{O}}(10^{-10})\lsim$ BR$(Z\to2\ell^P2\bar{\ell}^P)
\lsim {\cal{O}}(10^{-6})$, which is less than 
or comparable to the 
measured SM value \cite{PDG,Zto4l,smz4ln2}.
Note that $Z\to 2\ell2\tau,\,4\tau$ 
decays are experimentally unconstrained to date.

Let us finally remark that $\la\sim 0.1$
and $\mu\gsim 100$ GeV imply $\nu^c$
$\approx 333$ GeV. Hence, 
$0.1~{\rm GeV}$ $\lsim$ $|m_{\n_{i+3}}|$ $\lsim$ $45$ GeV predicts 
$0.00015$ $\lsim$ $\ka_i$ $\lsim$ $0.07$. Similarly,
approximate formulas for $m_{\P_i}$ give
$0.07~{\rm GeV}$ $\lsim$ $|{\aka}_i|$ $\lsim$ $29$ GeV.
The relative sign difference between $\ka$ and $\aka$
predicts $0.08~{\rm GeV}$ $\lsim$ $m_{\S_i}$ $\lsim 37$ GeV,
which does not introduce a large error in the $m_{\S_i}\sim
m_{\P_i}$ assumption. Here we have used 
$m^2_{\S_i}\approx m^2_{\n_{i+3}}+(\ka\aka)_i\nu^c$, 
$m^2_{\P_i}\approx -3(\ka\aka)_i\nu^c$, 
and $m_{\n_{i+3}} \approx 2\ka_i\nu^c$ 
in the limit of vanishingly small $\la$.

Concerning $\w$, leading decay widths are approximately given by
{\small
\bea\label{W-approx}
\Gamma^1_{\w\C\n} &\approx&\frac{9g^2_2 Y^2_\nu v^2_u M_W}{144\pi\mu^2}
\times\left(1-\frac{m^2_{\n_{j+3}}}{M^2_W}\right),\nonumber\\
\Gamma^2_{\w\C\n} &\approx&\frac{9 g^2_2 \lambda^2 Y^2_l 
v^2_u {\nu^2} M_W}{16\pi\mu^4}
\times\left(1-\frac{m^2_{\n_{j+3}}}{M^2_W}\right),
\eea}
where we have skipped terms like $m^2_{\c_i}/M^2_W$ due to their smallness
and the factor $``9$'' in the numerator appears after summing
over all possible combinations.
Now, as before, $\mu\approx {\cal{O}}(v)$,
$\la=0.1$, and $M_W=80.385$ GeV \cite{PDG} with  
a light $\n_{j+3}$ and a maximum of
$(Y_\nu,\,\nu)$ $\sim (10^{-6},\,10^{-4}$ GeV) 
for a TeV-scale seesaw \cite{mn1,mn2,PG1,PG2,Hirsch2009,mnSCPV}
give a maximum total new $\Gamma_{\w}$ $\sim {\mathcal{O}}(10^{-9}$ MeV) 
$\ll 34$ MeV, as evaluated 
in Sec. III. This new $\Gamma_{\w}$ decreases 
further with larger $\mu$ values. Note that unlike
the new $Z$ decays, we do not consider extra
factors coming from the field decomposition (see Ref. \cite{PG5})
which would produce further suppression.
Here we assume ${Y_\ell}$ ${\sim {\cal{O}} (1)}$ 
(which holds true for $\tb\gg1$)
\textit{only} for a maximum estimate. 
Hence, together with BR$(\n_{i+3}\to x^D\bar{x}^D+\MET)\approx 1$, 
one gets BR$(\w\to{{\ell}^\pm}^P/{\tau^\pm}^P$ $+~x^D\bar{x}^D$ $+~\MET)$ 
$\sim 3\times10^{-13}$ as a maximum.
This BR can at most reach ${\mathcal{O}}(10^{-12})$ when
$\mu\approx 100$ GeV.

\vspace*{0.15cm}
{\section*{VII. PROBING NEW {\boldmath $Z,\,\w$} 
DECAYS AT COLLIDERS}}


Production cross sections for $\w$ and $Z$
at the LHC run-I (center-of-mass energy E$_{\rm CM}=8$ TeV) 
are estimated as $\sigma_{\rm prod}$ $\approx 8.6\times 10^4$ 
and $2.5\times 10^4$ pb, respectively. {\textsc{pythia}~(version 6.409)} 
\cite{pythia} has been used for this purpose
(verified with \textsc{madgraph5} version 1.4.2 \cite{madgraph}),
with leading-order {\textsc{cteq6l1}} 
parton distribution functions \cite{pdf} having 
initial- and final-state radiations 
and multiple interactions switched on. 
For run-II, i.e., E$_{\rm CM}$ $=13$ and $14$ TeV,
$\sigma_{\rm prod}$ scales by a factor of
$\sim 2$.

The huge BR suppression for new $\w$ decays 
predicts $\approx 0.15$ events 
[${\mathcal{O}}(1)$ events with BR$\sim {\mathcal{O}}
(10^{-12})$ at the parton level],
taking E$_{\rm CM}=14$ TeV and an
integrated luminosity ${\mathcal{L}}=3000~{\rm fb}^{-1}$. 
Clearly, even without further reduction from the efficiency
issues, these practically background-free signals are
rather difficult to detect with both current and 
upcoming collider searches (e.g., 
\textit{MegaW} and \textit{OkuW} modes of the Linear
Collider \cite{lc} and Triple Large Electron-Positron Collider
(TLEP) \cite{tlep} with
about $2\times10^6$ and $7\times10^8$ $\w$ bosons/year, respectively).
Detecting these signals may appear feasible 
by using the techniques from the flavor sector,
especially when BR$(B^0_s\to\mu^+\mu^-)$
and BR$(\mu \to e \gamma)$ have already been probed
up to ${\cal{O}}(10^{-10})$ \cite{smallbr}
and ${\cal{O}}(10^{-13})$ \cite{meg2}, respectively.
The latter would reach ${\cal{O}}(10^{-19})$ in the
near future \cite{lfvf}.

On the other hand, unusual $Z$ decays 
(both prompt and displaced) for the LHC run-I with $\mathcal{L}=25$ fb$^{-1}$
gives about 62500 parton-level events with the
maximum BR estimate. Note that, as already stated, not all
of the daughter particles are hard, i.e., with high transverse momentum.
Hence, this number will reduce
further considering practical issues like $b$-tagging ($\tau$-tagging)
efficiency \cite{tdr-btau}, faking, etc. For displaced decays, the reconstruction
efficiency of the displaced vertices diminishes this number
even further. Assuming an optimal scenario with
a detection efficiency of 50\% (25\%) for the two leading (subleading) 
daughters, experimentally one 
can detect only $\approx$ 1.5\% of the parton-level events, i.e., 
about 938 surviving events. Repeating the same exercise
for the LHC run-II with E$_{\rm CM}=13$ TeV and $\mathcal{L}=100$ fb$^{-1}$,
one gets about 7500 surviving events out of $5\times10^5$ parton-level 
events. Making a similar analysis for $Z\to 4b^P$ in the SM with a BR of
$3.6^{+1.3}_{-1.3}\times 10^{-4}$ gives $3375^{+1219}_{-1219}$
and $27000^{+9750}_{-9750}$ events for run-I and run-II, respectively.

It is now apparent
that the final number of events for these spectacular signatures
remains below that coming from the errors of the SM measurement.
Thus, unless one adopts dedicated experimental searches
(as already mentioned) to perform a faithful construction 
of the $2b$ jets $m_{\rm inv}$ (${\rm M_{T2}}$ for $\tau$ jets), 
which is supposed to peak around $m_{\S_i,\,\P_i}$, 
it remains hardly possible to
isolate these new signals from the SM backgrounds. 
Consequently, the region of parameter space
with $2m_b\lsim m_{\S_i,\,\P_i}\lsim M_Z/2$ remains
unconstrained from the existing experimental results. 
A similar logic remains applicable for the $Z\to 4\ell^P$
process, through leptonic $\tau$ decays following
$Z\to4\tau^P$. Here, for run-I, and $\mathcal{L}=20.3$ fb$^{-1}$ 
as studied by the ATLAS Collaboration \cite{smz4ln2}, one gets a
maximum of 761 events at the parton level, including $e,\mu$ flavors,
while the SM number is $1624^{+142}_{-142}$ with the latest BR$(Z\to 4\ell^P)$ \cite{smz4ln2}.
So, unless $\mu,\,A_\la,\,\widetilde m\approx {\mathcal{O}}(v)$, 
these novel signals normally lie beneath their SM counterparts,
and hence the region of parameter space
giving $2m_\tau$ $\lsim$ $m_{\S_i,\,\P_i}$ $\lsim$ $2m_b$
remains exempted from the experimental constraints.
The existing results for BR$(Z\to 4\ell^P)$ measurements in
the SM \cite{PDG,Zto4l,smz4ln2}, however, put constraints
on the corner of parameter space with $m_{\S_i,\,\P_i}\lsim 2m_\tau$,
unless $A_\la,\,\widetilde m,\,\mu > v$ or $\la \ll 0.1$.
The $m_{\S_i,\,\P_i}\lsim 2m_\tau$ region, 
however, is already challenged by a class of flavor
observables \cite{Upsilon}. 

Note that, concerning statistics, the upcoming colliders 
(\textit{GigaZ} and \textit{TeraZ} modes
of the Linear Collider \cite{lc} and TLEP \cite{tlep} 
with about $2\times10^9$ and $7\times10^{11}$ $Z$ bosons/year, respectively)
are comparable to the LHC run-II, e.g., with \textit{TeraZ}
one expects about $10^4$ to $10^8$ novel $Z$-decay events when
the BR varies from $10^{-8}$ to $10^{-4}$. Nonetheless,
the unprecedented accuracy of these new colliders would
proficiently constrain the concerned region of parameter 
space giving rare signals, e.g., in TLEP the error
in $\Gamma_Z$ would reduce to $<$ 10 keV \cite{tlep}.

\vspace*{0.25cm}
{\section*{VIII. CONCLUSIONS}}

In conclusion, we have presented a complete analytical study of all 
possible two-body nonstandard
$\w$ and $Z$ decays in the context of SUSY theories.
The detection of these decays at the LHC or in a 
future collider (of course, with evolved search
criteria) would provide an unambiguous sign of new physics
beyond the SM. The presence of these signals 
will also offer an experimental test for the $\mn$, 
the \textit{simplest} extension of the MSSM to accommodate 
these unique signatures, apart from solving 
shortcomings of the MSSM. We also show that other variants of SUSY models
can be investigated in a similar way. After the latest
and improved measurement of $Z\to 4\ell^P$ processes by the ATLAS
Collaboration, we look forward to further analysis
in this direction, both from the CMS and
ATLAS Collaborations, involving $b$ jets or $\tau$ jets, coming
from both prompt and displaced origin. Missing evidence
of these modes, in the presence of a proper experimental setup,
would constrain the feasibility of light $\S_i,\P_i$, and $\n_{i+3}$
states. Even the existing bound on the branching fractions
can disfavor certain regions of the parameter space,
e.g. $\la$ around $0.1$ with $\mu\sim 100$ GeV. 
We further observed that the region of parameter
space with $\mu,\,\ala,\,\widetilde m \approx {\cal{O}}(v)$
and $m_{\S_i,\,\P_i}\lsim 2m_\tau$
is largely excluded from $Z\to 4\ell^P$ searches
in the SM unless one considers $\la \ll 0.1$.

\vspace*{0.5cm}

\vspace*{0.15cm}
{\section*{ACKNOWLEDGEMENTS}}

The work  of P.G. and C.M. is supported in part 
by the Spanish MINECO  under grant 
FPA2012-34694 and under the ``Centro de Excelencia 
Severo Ochoa'' Programme SEV-2012-0249, and by the Comunidad 
de Madrid under grant HEPHACOS 
S2009/ESP-1473. The work of D.L. is supported by the Argentinian CONICET. 
V.A.M. acknowledges support by the Spanish MINECO under the project FPA2012-39055-C02-01, 
by the Generalitat Valenciana through the project PROMETEO~II/2013-017 and by the 
Spanish National Research Council (CSIC) under the JAE-Doc program co-funded by 
the European Social Fund (ESF). The work of R.R. is  supported 
by the Ram\'on y Cajal program of the Spanish MINECO and also thanks the support of the 
MINECO under grant FPA2011-29678. The authors also acknowledge the support of the 
MINECO's Consolider-Ingenio  2010 Programme under grant MultiDark CSD2009-00064.






\end{document}